# Delay-Optimal Power and Precoder Adaptation for Multi-stream MIMO Systems

Vincent K. N. Lau and Yan Chen

*Abstract*—In this paper, we consider delay-optimal MIMO precoder and power allocation design for a MIMO Link in wireless fading channels. There are $L$ data streams spatially multiplexed onto the MIMO link with heterogeneous packet arrivals and delay requirements. The transmitter is assumed to have knowledge of the channel state information (CSI) as well as the joint queue state information (QSI) of the $L$ buffers. Using $L$-dimensional Markov Decision Process (MDP), we obtain optimal precoding and power allocation policies for general delay regime, which consists of an online solution and an offline solution. The online solution has negligible complexity but the offline solution has worst case complexity $\mathcal{O}((N+1)^L)$ where $N$ is the buffer size. Using *static sorting* of the $L$ eigenchannels, we decompose the MDP into $L$ independent 1-dimensional subproblems and obtained low complexity offline solution with linear complexity order $\mathcal{O}(NL)$ and close-to-optimal performance.

## I. INTRODUCTION

Multiple Input Multiple Output (MIMO) communication is well-known to boost the wireless spectral efficiency through spatial multiplexing. Substantial performance gain could be obtained by power and precoder adaptation according to the channel state information is available at the transmitter (CSIT). In [1], [2], a linear MIMO precoder design framework is proposed to minimize the weighted sum of mean square errors (MSE) assuming knowledge of perfect CSIT. In [3] and [4], MIMO precoder design utilizing either limited feedback or outdated CSIT is proposed. Yet, all these works assumed that the transmitter has infinite buffer and the information flow is delay insensitive, and focused on optimizing the physical layer performance (such as capacity, throughput or MSE). In practice, it is very important to consider the delay performance in addition to the conventional physical layer performance in MIMO transceiver design.

A combined framework taking into account of both queueing delay and physical layer performance is not trivial as it involves both the queueing theory (to model the queue dynamics) and information theory (to model the physical layer dynamics). In [5], it is shown that naive water-filling (which is optimal in information theoretical sense) is not always a good strategy with respect to the delay performance. In general, there are two approaches to deal with delay problems. The first approach converts the delay constraint into average rate constraint using tail probability at large delay regime and solve the optimization problem using information theoretical formulation based on the rate constraint [6]–[8]. While this approach allows potentially simple solution, the control policy will be a function of CSIT only and such control will be good only for large delay regime. In general, the delay-optimal power and precoder adaptation will be a function of both the CSI and the queue state information (QSI). In the second approach, the problem of finding the optimal control policy (to minimize delay) is cast into a *Markov Decision Problem* (MDP) or stochastic control problem [9]. Unfortunately, it is well-known that there is no easy solution (e.g. value iteration and policy iteration) to MDP in general, even for the simple scenario like SISO channel [10], [11]. In [12], [13], the authors showed that the longest queue highest possible rate (LQHPR) policy is delay-optimal for symmetric multi-access fading channels. Works considering delay sensitive scheduling can be found in [14] and [15]. While all the above works addressed different aspects of the delay sensitive resource allocation problem, there are still some first order issues to be addressed.

- **Low complexity optimal control policy for delay sensitive resource allocation problem in general delay regime** Most of the existing works considered large delay asymptotic solutions. However, practical operating region for delay sensitive traffics are usually on the low delay regime and the asymptotic simplifications cannot be applied. Hence, it is important to obtain low complexity control policy for general delay regime.
- **Coupling among multiple delay-sensitive heterogeneous data streams** Most of the above works considered single stream wireless link only [16]. While [12], [13] considered multi-user systems, the framework applies to situations with symmetric (homogeneous users) only and cannot be extended to situations with heterogeneous users. When we have heterogeneous data streams, the problem will be difficult as the optimal policy will generally be coupled with the joint queue state of all the heterogeneous streams. The general solution involves solving multi-dimensional MDP with exponential order of complexity w.r.t. the number of streams.

In this paper, we shall attempt to address the above issues for the delay-sensitive multi-stream MIMO power and precoder adaptation design. Specifically, we consider an $N_t \times N_r$ MIMO link with $L \leq \min\{N_t, N_r\}$ spatially multiplexed heterogeneous data streams (with different delay requirements). This represents an important scenario where a multi-antenna terminal receiving data from multiple application streams (with different delay requirements) simultaneously through a MIMO link from the base station. The transmitter is assumed to have knowledge of both the CSI and the QSI. Using $L$-

This paper is funded by RGC 615407.

Vincent K. N. Lau (e-mail: eeknlau@ee.ust.hk) is with Department of Electronic and Computer Engineering, Hong Kong University of Science and Technology (HKUST), Clear Water Bay, Kowloon, Hong Kong. Yan Chen (e-mail: {qiupl418}@zju.edu.cn) is with Institute of Information and Communication Engineering, Zhejiang University, Hangzhou, China.

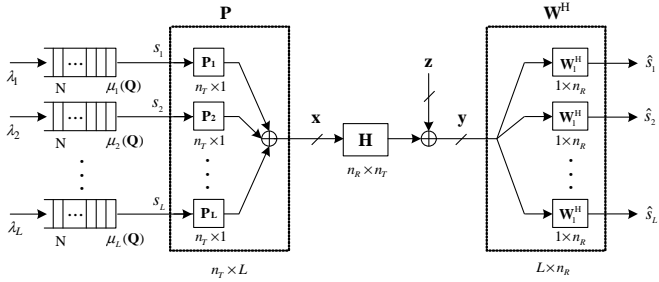

Fig. 1. Top level system model.

dimensional MDP formulation on the embedded markov chain, we derive an optimal control policy to minimize the weighted average delays of the $L$ application streams for general delay regime. The optimal policy consists of an online procedure and an offline procedure. The online procedure has negligible complexity but the offline procedure could be quite complex for large $L$. Using *static eigenchannel mapping*, we decompose the $L$-dimensional MDP problems into $L$ one-dimensional MDP subproblems and obtain a low complexity solution with worst case complexity of $\mathcal{O}(NL)$ but close-to-optimal performance.

The paper is organized as follows. In Section II, we shall elaborate the system model, physical layer model as well as the queue model. In Section III, we formulate the delay-sensitive precoder and power adaptation design as an MDP. In Section IV, we derive the low complexity optimal control policy. In Section V, we elaborate the extensions when the CSIT is outdated. Section VI illustrates the delay performance of the proposed algorithm by simulations. Finally, we conclude with a brief summary of results in Section VII.

## II. SYSTEM MODELS

In this section, we shall elaborate the system model, physical layer model as well as the underlying queueing model. Fig. 1 illustrates the top level system model where $L$ application streams are spatially multiplexed and delivered to a multi-antenna terminal (with $N_r$ antennas) from a multi-antenna source (with $N_t$ antennas). These $L$ application streams may have different source arrival rates and delay requirements[1].

### A. MIMO Physical Layer Model

We consider the use of MIMO linear transceivers, composed of a linear precoder at the transmitter (represented by a matrix $\mathbf{P} \in \mathbb{C}^{N_t \times L}$) and a linear equalizer at the receiver (represented by a matrix $\mathbf{W} \in \mathbb{C}^{N_r \times L}$). The transmitted vector $\mathbf{x} \in \mathbb{C}^{N_t}$ is given by $\mathbf{x} = \mathbf{P}\mathbf{s}$ where $\mathbf{s} \in \mathbb{C}^L$ is the normalized data symbols from the $L$ application streams with $\mathbb{E}\left(\mathbf{s}\mathbf{s}^H\right) = \mathbf{I}$, and the total average transmitted power should satisfy $\mathbb{E}\left[\|\mathbf{x}\|^2\right] = \text{Tr}\left(\mathbf{P}\mathbf{P}^H\right)$. Similarly, the estimated received symbols (corresponding to the equalizer outputs) is given by $\widehat{\mathbf{s}} = \mathbf{W}^H\mathbf{y}$ where $\mathbf{y} \in \mathbb{C}^{N_r}$ is the channel outputs, i.e. $\mathbf{y} = \mathbf{H}\mathbf{x} + \mathbf{z}$. Here $\mathbf{H} \in \mathbb{C}^{N_r \times N_t}$ is the MIMO channel state information (CSI) and $\mathbf{z} \in \mathbb{C}^{N_r}$ is a zero-mean circularly symmetric complex Gaussian noise vector with normalized covariance $\mathbf{I}$. Both the transmitter and the receiver are assumed to have perfect knowledge of the MIMO CSI $\mathbf{H}$[2].

As a result, the equivalent channel (with precoder, MIMO channel and the equalizer) for the $L$ data stream is $\widehat{\mathbf{s}} = \mathbf{W}^H\mathbf{H}\mathbf{P}\mathbf{s} + \mathbf{W}^H\mathbf{z}$ and the SINR of the i-th data stream is $SINR_i(\mathbf{P}) = |\mathbf{w}_i^H\mathbf{H}\mathbf{p}_i|^2/\mathbf{w}_i^H\mathbf{A}_i\mathbf{w}_i$, where $\mathbf{A}_i = \sum_{j \neq i}\mathbf{H}\mathbf{p}_j\mathbf{p}_j^H\mathbf{H}^H + \mathbf{I}$ and $\{\mathbf{p}_i\}$ denotes the i-th column of the precoding matrix $\mathbf{P}$. For sufficiently high SINR, the symbol error probability (SEP) of QAM constellation is [17]:

$$P_e(\mathbf{H}) \leq \kappa_1 Q\left(\sqrt{\frac{3SINR_i}{2^{R_i}-1}}\right) \leq \frac{\kappa_1}{2}\exp\left(\frac{3SINR_i}{2(2^{R_i}-1)}\right)$$

for some constant $\kappa_1$. Hence, given a sufficiently small target SEP $\epsilon$, the data rate $R_i$ (bits per symbol) of the i-th data stream is related to the $SINR_i(\mathbf{P})$ as $R_i = \log_2(1 + \alpha(\epsilon)SINR_i(\mathbf{P}))$, where $\alpha(\epsilon)$ is some constant depending on the target SEP $\epsilon$. Since the receiver has perfect CSIR and the data rate is an increasing function of $SINR_i$, it is shown that for any precoder $\mathbf{P}$, *Wiener filter* $\mathbf{W} = (\mathbf{H}\mathbf{P}\mathbf{P}^H\mathbf{H}^H + \mathbf{I})^{-1}\mathbf{H}\mathbf{P}$ can simultaneously maximize $\{SINR_1, .., SINR_L\}$ [18]. As a result, the conditional average SINR of the i-th data stream after Wiener filtering is given by $SINR_i(\mathbf{P}) = \mathbf{p}_i^H\mathbf{H}^H\mathbf{A}_i^{-1}\mathbf{H}\mathbf{p}_i$. Define the *instantaneous MSE* matrix as

$$\mathbf{E}(\mathbf{P}) = \mathbb{E}\left[(\widehat{\mathbf{s}} - \mathbf{s})(\widehat{\mathbf{s}} - \mathbf{s})^H\right] = (\mathbf{I} + \mathbf{P}^H\mathbf{H}^H\mathbf{H}\mathbf{P})^{-1}. \quad (1)$$

Note that the diagonal elements of $\mathbf{E}$ contains the instantaneous MSEs of the $L$ data streams. Using matrix inversion lemma [19], it can be shown that $SINR_i(\mathbf{P}) = \mathbf{E}_{ii}^{-1}(\mathbf{P}) - 1$. Hence, the supported data rate at the target SEP $\epsilon$ given by $R_i = \log_2\left(1 + \alpha(\epsilon)(\mathbf{E}_{ii}^{-1}(\mathbf{P}) - 1)\right)$.

### B. Queue Model, System States and Control Policy

In this paper, the time dimension is partitioned into *scheduling slots* (each slot has $\tau$ channel uses) and we assume that the CSI $\mathbf{H}$ remains quasi-static[3] within a scheduling slot and i.i.d. between scheduling slots. There are $L$ buffers (each of length $N$) at the transmitter for the $L$ application streams respectively. For simplicity, we assume the $L$ application sources follow Poisson arrival with mean arrival rates $(\lambda_1, .., \lambda_L)$ (number of packets per channel use). The packet length of the $l$-th data source, $N_l$, follows exponential distribution with mean packet size $\overline{N_l}$ (bits per packet). The transmitter is assumed to have knowledge of the QSI of the $L$ buffers. Specifically, the QSI at time $t$ is denoted by $\mathbf{Q}(t) = (Q_1(t), .., Q_L(t) \in \{0, .., N\}^L)$ where $Q_l(t)$ is the number of packets in the $l - th$ buffer at time $t$. As a result, the *observed system state* at the transmitter, $\chi = (\mathbf{H}, \mathbf{Q})$, consists of both the CSIT and the joint QSI. Given an observed system state realization $\chi$, the transmitter may adjust the transmit power and precoding matrix $\mathbf{P}$ according to a *stationary precoding policy*[4] $\pi = \{\mathbf{P}(\chi)\}$ defined below.

---

[1]This corresponds to the scenario where the multi-antenna terminal may be running different applications simultaneously.

[2]We elaborate the case when the CSIT is outdated in section V.

[3]This assumption is realistic for pedestrian mobility users where the channel coherence time is around 50 ms but typical frame duration is less than 5ms in next generation wireless systems such as WiMAX.

*Definition 1:* (**Stationary Precoding and Power Control Policy**) A stationary transmit power and precoding policy $\pi : \{0, .., N\}^L \times \mathbb{C}^{N_r \times N_t} \to \mathbb{C}^{N_t \times L}$ is defined as the mapping from the currently observed system state $\chi = (\mathbf{Q}, \mathbf{H})$ to a linear precoder $\pi(\chi) = \mathbf{P}(\chi)^5$. The set of all feasible stationary policies is defined as $\mathcal{P} = \{\pi : \operatorname{Tr}(\mathbb{E}\left[\pi(\chi)\pi^H(\chi)|\mathbf{Q}\right]) > 0, \forall \mathbf{Q} \in \{0, 1, 2, ..., N\}^L\}$.

Since the packet length is exponentially distributed with mean packet length $\overline{N_i}$, the packet service time follows exponential distribution with conditional mean service rate (conditioned on system state $\chi$) [packets per channel use]

$$\mu_i(\chi) = \frac{1}{\overline{N_i}} \log_2\left(1 + \alpha(\epsilon)(\mathbf{E}_{ii}^{-1} - 1)\right). \quad (2)$$

The overall delay dynamics of the $L$-stream multiplexed MIMO system can be modeled by $L$ $M/M/1$ queues as illustrated in Fig. 1. The $L$ queues are coupled together via the precoding policy $\mathbf{P}$ and the transmit power constraint. We shall derive an optimal stationary precoding policy to minimize the average delays of the $L$ spatially multiplexed data streams subject to average transmit power constraint. Specifically, the average delay (in packets) of the $i$-th data stream is given by

$$\overline{T}_i(\pi) \triangleq \limsup_M \frac{1}{M} \mathbb{E}\left[\sum_{m=1}^M Q_{i,m}\right], \quad \forall i \in \{1, ..., L\} \quad (3)$$

where $Q_{i,m} = Q_i(m\tau)$ is the QSI of the $i$-th buffer observed at $t = m\tau$. The average transmit power constraint is given by:

$$\overline{P_{tx}}(\pi) \triangleq \limsup_M \frac{1}{M} \mathbb{E}\left[\sum_{m=1}^M \operatorname{Tr}(\pi(\chi_m)\pi^H(\chi_m))\right] \le P_0 \quad (4)$$

where $\pi(\chi_m)$ denotes the precoder applied at $t = m\tau$. Note that the transmitter may adjust the precoding and power control actions only at the beginning of scheduling slots and the control action remains unchanged in between the scheduling slots. The average delay is related to the transmit power via the packet service rates $\{\mu_1(\chi), ..., \mu_L(\chi)\}$. The delay optimization problem can be formally written as:

*Problem 1:* (**Delay Optimal Policy**) For some $\beta = (\beta_1, \beta_2, ..., \beta_L)$ (such that $\beta_i > 0$ for all $i$), we seek to find a stationary policy $\pi \in \mathcal{P}$ that minimizes

$$J_\beta^\pi(\chi_0) = \sum_{i=1}^L \beta_i \overline{T}_i(\pi) + \gamma \overline{P_{tx}}(\pi). \quad (5)$$

where $\chi_0$ denotes the initial system state. The positive weighting factors $\beta$ indicate the relative importance of buffer delay among the $L$ data streams and for each given $\beta$, the solution to (5) corresponds to a point on the Pareto optimal delay tradeoff boundary. The constant $\gamma > 0$ is the Lagrange multiplier for the average transmit power constraint in (4).

---

[4] It is shown [9] that for finite state MDP, stationary and history independent policy is optimal. Hence, there is no loss of generality to consider policy that is function of current system state only.

[5] Note that since a linear precoder $\mathbf{P}$ can be decomposed into $\mathbf{U}_P \mathbf{\Sigma}_P \mathbf{V}_P$ where $\mathbf{U}_P$ and $\mathbf{V}_P$ are unitary matrices (denoting the precoding actions) and $\mathbf{\Sigma}_P$ is a diagonal matrix (denoting the power allocation action), we shall represent both the precoding and power allocation actions by a single precoding matrix $\pi(\chi)$.

## III. Markov Decision Problem Formulation

In this section, we shall formulate the delay minimization problem as Markov Decision Process and discuss the optimality condition. We shall first introduce the embedded Markov chain and the induced reward random sequence.

### A. Embedded Markov Chain and MDP Formulation

Recall that $\{\mathbf{Q}(t)\}$ is the continuous time random process (denoting the joint queue state of the $L$ data streams) and $\{\mathbf{Q}_m\}$ is the corresponding induced discrete time random process (denoting the joint queue states at observation epochs $\{0, \tau, 2\tau, ....\}$) with $\mathbf{Q}_m = \mathbf{Q}(m\tau)$. The problem of finding the optimal control policy $\pi$ to minimize system delay in Problem 1 is in general quite tedious even for obtaining numerical solutions. To obtain simple solution, we consider the case where the scheduling slot duration (or frame duration) $\tau$ is substantially smaller than the average packet interarrival time as well as average packet service time ($\tau \ll \frac{1}{\lambda}$ and $\tau \ll \frac{1}{\mu_i(\chi)}$)[6]. Suppose the system state at the $m$-th observation epoch is $\chi_m = \{\mathbf{H}_m, \mathbf{Q}_m\}$. At the $(m+1)$-th observation epoch $t = (m+1)\tau$, one of the following events may happen: 1) packet arrival from the $i$-th data source with probability $p_{q,q+1}^{(i)} = \lambda_i \tau$; 2) Packet departure from the $i$-th data buffer with probability $p_{q,q-1}^{(i)} = \overline{\mu_i}(\mathbf{Q}_m)\tau = \mathbb{E}_\mathbf{H}[\mu_i(\chi_m)|\mathbf{Q}_m]\tau$; 3) No change in the $i$-th buffer state with probability $p_{q,q}^{(i)} = 1 - p_{q,q-1}^{(i)} - p_{q,q+1}^{(i)}$.[7] Therefore, the *embedded discrete time random variables* $\{\mathbf{Q}_m\}$ is an *irreducible* Markov chain induced by a stationary policy $\pi \in \mathcal{P}$. In addition, given a stationary policy $\mathcal{P}$, the Markov chain $\{\mathbf{Q}_m\}$ depends on $\pi$ via the conditional average *packet service rate* $\overline{\mu_i}(\mathbf{Q}_m)$ only. On the other hand, since the CSIT $\{\mathbf{H}_m\}$ is i.i.d. between any two observation epochs, the optimization objective function (average cost per stage) $J_\beta^\pi(\chi_0)$ evaluated at the discrete time observation epochs can be expressed as:

$$J_\beta^\pi(\chi_0) = \limsup_M \frac{1}{M} \sum_{m=1}^M \mathbb{E}_\mathbf{Q}\left[g(\mathbf{Q}_m, \overline{\pi}(\mathbf{Q}_m))\right] \quad (6)$$

$$\text{where} \quad g(\mathbf{Q}_m, \overline{\pi}(\mathbf{Q}_m)) = \sum_{i=1}^L \beta_i Q_{i,m} + \gamma \operatorname{Tr}[\overline{\pi}(\mathbf{Q}_m)] \quad (7)$$

$$\overline{\pi}(\mathbf{Q}_m) = \mathbb{E}_\mathbb{H}\left[\pi(\chi_m)\pi^H(\chi_m)|\mathbf{Q}_m\right]. \quad (8)$$

Given a stationary policy $\pi$, the Markov chain $\{\mathbf{Q}_m\}$ induces a random sequence of reward functions $\{g(\mathbf{Q}_m, \overline{\pi}(\mathbf{Q}_m))\}$ depending on the chosen policy $\pi \in \mathcal{P}$. From (7), the evolution of the random sequence of reward function $\{g(\mathbf{Q}_m, \overline{\pi}(\mathbf{Q}_m))\}$ depends on $\pi \in \mathcal{P}$ via the conditional average transmit power cost $\operatorname{Tr}[\overline{\pi}(\mathbf{Q}_m)\overline{\pi}^H(\mathbf{Q}_m)]$ only. Hence, the delay-optimization problem in Problem 1 could be completely characterized by a multi-dimensional infinite horizon *Markov Decision Process* (MDP) with *partial system state* $\mathbf{Q}$, per-stage reward function $g(\mathbf{Q}, \overline{\pi}(\mathbf{Q}))$, and the conditional average precoding action $\overline{\pi}(\mathbf{Q})$. The state transition probability of the embedded MDP $\Pr[\mathbf{Q}_{m+1}|\mathbf{Q}_m, \overline{\pi}(\mathbf{Q}_m)]$ is illustrated in Fig. 2.

---

[6] This is a mild assumption which could be justified in many applications. For example, in WiMAX, a frame duration is around 2ms while the target queueing delay for video streaming is around 200ms or more.

[7] Since $\tau$ is small, the probability of multiple packet arrivals or departures among the $L$ data sources is negligible and hence $p_{q,p}^{(i)} = 0$ for $|p - q| > 1$.

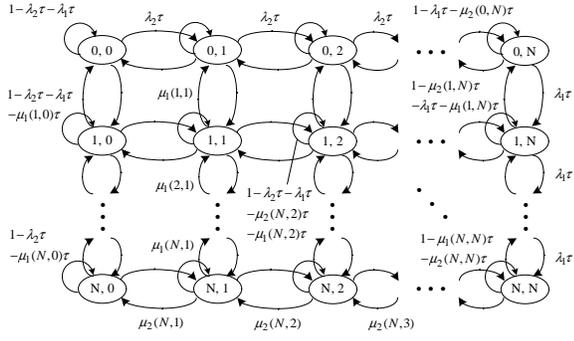

Fig. 2. State transition diagram for $L-$dimension Markov chain $\{\mathbf{Q_m}\}$ with $N$ states each dimension. $L = 2$ for illustration.

### B. Bellman Condition and Optimal Precoding Structure

In general, the sequence of average costs $\{\mathbb{E}[g(\mathbf{Q}_m, \overline{\pi}(\mathbf{Q}_m))]\}$ of the infinite horizon MDP under a chosen stationary policy $\pi \in \mathcal{P}$ may not converge at all. Since the induced Markov chain $\{\mathbf{Q}_m\}$ is irreducible for any stationary policy $\pi \in \mathcal{P}$, the limit of long run average cost $J_\beta^\pi(\chi_0)$ converges and is independent of the initial state $\chi_0$. For the infinite horizon MDP described by Fig. 2, the optimizing policy can be obtained by solving the *Bellman equation* [9] recursively w.r.t. $(\theta, \{V(q_1, .., q_L)\})$ as below:

$$\theta + V(q_1, .., q_L) = \inf_{\pi \in \mathcal{P}} \{g(q_1, ..., q_L, \overline{\pi}(q_1, ..., q_L))$$
$$+ \tau \sum_{i=1}^{L} \lambda_i V(q_1, .., (q_i + 1)_{\bigwedge N}, .., q_L)$$
$$+ \tau \sum_{i=1}^{L} \overline{\mu_i}(q_1, .., q_L) V(q_1, .., [q_i - 1]^+, .., q_L)$$
$$+ V(q_1, .., q_L) \left( 1 - \sum_{i=1}^{L} \tau \lambda_i - \sum_{i=1}^{L} \tau \overline{\mu_i}(q_1, .., q_L) \right) \} \quad (9)$$

for all $(q_1, .., q_L) \in \{0, 1, ..., N\}^L$ where $x_{\bigwedge y} = \min\{x, y\}$. If there is a $(\theta, \{V(q_1, .., q_L)\})$ satisfying (9), then $\theta = \inf_{\pi \in \mathcal{P}} J_\beta^\pi$ is the optimal average reward per stage. Furthermore, since the induced Markov chain $\{\mathbf{Q}_m\}$ is irreducible for any stationary policy $\pi \in \mathcal{P}$, the solution to (9) is unique.

Note that solution to (9) is still very complex due to the following. Firstly, it involves $L$-dimensional recursions and as a result, brute-force solutions have exponential order of complexity w.r.t. $L$. Secondly, each step of the recursion involves optimization w.r.t. matrix precoder $\pi(\chi)$. In the following, we shall utilize the underlying structure to deduce the optimal precoding structure for $\pi(\chi)$ first.

Given any QSI $\mathbf{Q}$ and $V(\mathbf{Q})$, let $\overline{\pi}(\mathbf{Q}) = \{\mathbf{P} = \pi(\mathbf{Q}, \mathbf{H}) \in \mathbb{C}^{N_t \times L} : \forall \ \mathbf{H} \in \mathbb{C}^{N_r \times N_t}\}$ be the set of all precoding actions per any possible CSIT realization (given a certain QSI realization $\mathbf{Q}$). The optimization in the RHS of (9) is equivalent to the following form:

$$\min_{\overline{\pi}(\mathbf{Q})} \mathbb{E}_{\mathbf{H}} \left[ \mathcal{G} \Big( \mathbf{d} \big( \mathbf{E}(\mathbf{P}) \big), \mathbf{P} \Big) \right] \quad (10)$$

where $\mathbf{d}(\mathbf{A})$ denotes the diagonal elements of matrix $\mathbf{A}$ and

$$\mathcal{G}(d_1, .., d_L, \mathbf{P}) =$$
$$\frac{\tau}{N} \sum_{i=1}^{L} \left( V(q_1, .., [q_i - 1]^+, .. q_L) - V(q_1, .., q_L) \right)$$
$$\times \log_2 \left( 1 + \alpha(\epsilon)(d_i^{-1}(\mathbf{E}(\mathbf{P})) - 1) \right) + \gamma \text{Tr}[\mathbf{PP}^H]. \quad (11)$$

Note that $\mathbf{d}(\mathbf{E}(\mathbf{P}))$ is related to the precoding matrix $\pi(\mathbf{H}, \mathbf{Q})$ according to (1) and (2). Since the optimization variables in (11) involve the set of actions $\overline{\pi}(\mathbf{Q})$ for all CSIT realizations, the problem can be decomposed into solving $\min_{\mathbf{P}} \mathcal{G}(\mathbf{d}(\mathbf{E}(\mathbf{P})), \mathbf{P})$ for each CSIT and QSI realization. To derive the optimal solution, we first have the following lemma.

*Lemma 1:* If $(\theta, \{V(q_1, .., q_L)\})$ is a solution to the Bellman equation (9), then $V(q_1, .., q_L)$ is a monotonically non-decreasing function in all the $L$ arguments.

As a result of Lemma 1, $\mathcal{G}(d_1, .., d_L; \mathbf{P})$ is a Schur-concave function in $(d_1, .., d_L)$ and the optimal transmit precoder matrix is summarized in the following theorem.

*Theorem 1:* (**Optimal Precoding Matrix**) For any realization of system state $\chi$(QSI $\mathbf{Q}$, CSIT $\mathbf{H}$), the optimal precoding action $\pi(\chi) = \mathbf{P}$ w.r.t. (10) is given by:

$$\pi(\chi) = \mathbf{P} = \mathbf{U}\mathbf{\Sigma}_p \quad (12)$$

where $\mathbf{U} \in \mathbb{C}^{N_t \times L}$ is a unitary matrix consisting of $L$ eigenvectors of $\mathbf{H}^H \mathbf{H}$ corresponding to the $L$ largest eigenvalues and $\mathbf{\Sigma}_p = diag\{\sqrt{p_1}, \ldots, \sqrt{p_L}\}$ is a diagonal matrix containing the power allocations over the $L$ spatial channels. Note that the $L$ largest eigenvalues $\{\xi_1, .., \xi_L\}$ are sorted in the same order as $\eta_i = V(q_1, \ldots, q_L) - V(q_1, \ldots, [q_i - 1]^+, \ldots, q_L)$.

All the proofs are omitted for lack of space and interested readers can refer to our full version in [20] for details. In general, the delay-optimal precoding and power allocation actions should be a function of both CSIT and QSI. From Theorem 1, the optimal precoding matrix $\mathbf{U}$ seems to be a function of CSIT $\mathbf{H}$ only. However, this is not the case as the ordering of the $L$ largest eigenvalues $\{\xi_1, .., \xi_L\}$ has to be sorted in the same order as $\{\eta_i\}$ where $\eta_i = V(q_1, .., q_L) - V(q_1, .., [q_i - 1]^+, .. q_L)$ is a function of the QSI $\mathbf{Q}$. Hence, the precoding matrix $\mathbf{U}$ is indeed a function of both the CSIT and QSI (implicitly) and it's because of this sorting requirement of the eigenvalues that makes the MDP analysis of the $L$ data streams coupled together.

*Remark 1:* Note that from Theorem 1, the delay-optimal precoder has the MIMO-channel diagonalizing structure and as a result, the subsequent delay-optimization and solutions can be applied to general L-parallel channels such as OFDM systems as well.

### C. Optimal Power Allocation Policy

Using the precoder structure given by Theorem 1, the conditional average MSE becomes $\mathbf{d}[\mathbf{E}] = [(1 + p_1\xi_1)^{-1}, \ldots, (1 + p_L\xi_L)^{-1}]$. Hence, the conditional average service rate becomes

$$\overline{\mu_i}(\mathbf{Q}) = \frac{1}{N_i} \log_2 \left( 1 + \alpha(\epsilon) p_i \xi_i \right). \quad (13)$$

Therefore, without loss of generality, we shall consider optimization w.r.t. the power allocation policy $\varphi$ defined as:

*Definition 2:* (**Power Allocation Policy**) A power allocation policy $\varphi : \{0,..,N\}^L \times \mathbb{C}^{N_r \times N_t} \rightarrow \mathbb{R}_+^L$ is defined as the mapping from the currently observed system state $\chi = (\mathbf{Q}, \mathbf{H})$ to a power allocation vector $\varphi(\chi) = (p_1,..,p_L)$ where $\varphi_i(\chi) = p_i$ gives the power allocation to the $i$-th data stream. Furthermore, $\varphi(q_1,..,q_L) = \{(p_1,..,p_L) = \varphi(\mathbf{H}, \mathbf{Q} = (q_1,..,q_L)) : \mathbf{H} \in \mathbb{C}^{N_r \times N_t}\}$ denotes the set of power allocation actions for all CSIT realizations at a given QSI $\mathbf{Q} = (q_1,...,q_L)$.

The Bellman equation in (9) can thus be written as:

$$\sum_{i=1}^{L} \lambda_i \delta V_i(q_1,..,(q_i+1)_{\Lambda N},..,q_L) + \sum_{i=1}^{L} \beta_i q_i$$
$$-\phi(\delta V_1(q_1,..,q_L),...,\delta V_L(q_1,..,q_L) = \theta, \quad (14)$$

for $q_i = 0,..,N$ (and the initial condition can be set as $V(0,..,0) = 0$). $\delta V_i(q_1,..,q_L) \overset{\Delta}{=} \tau(V(q_1,..,q_i,..,q_L) - V(q_1,..,[q_i-1]^+,..,q_L))$, and $\phi(\eta_1,...,\eta_L)$ is defined as

$$\sup \mathbb{E}_{\mathbf{H}} \left[ \sum_{i=1}^{L} \left( \frac{\eta_i}{N_i} \log_2 \left( 1 + \alpha(\epsilon) p_i(\mathbf{H}) \xi_{[i]} \right) - \gamma p_i(\mathbf{H}) \right) \right],$$

The supremum is taken w.r.t $p_1(\mathbf{H}),\ldots,p_L(\mathbf{H})$ and $\{\xi_{[i]}\}$ denotes the $L$ largest eigenvalues of $\mathbf{H}^H \mathbf{H}$ sorted in the same order as $\{\eta_1,..,\eta_L\}$. Using standard optimization technique, the optimizing power allocation policy for $\phi(\eta_1,...,\eta_L)$ is given by the standard water-filling solution:

$$p_i^*(\mathbf{H}, \eta_1,..,\eta_L) = \left( \frac{\eta_i}{\overline{N_i}\gamma} - \frac{1}{\alpha(\epsilon)\xi_{[i]}} \right)^+. \quad (15)$$

Hence, the Bellman equation in (14) can be solved using *policy iteration* [9] in an offline manner. Once the solution of the Bellman equation in (14) is determined, the optimal power allocation (given a CSIT and QSI realization) is given by $\varphi_i^*(\mathbf{H}, \mathbf{Q}) = p_i^*(\mathbf{H}, \delta V_1(\mathbf{Q}),..., \delta V_L(\mathbf{Q}))$ as defined in (15). Using the optimal power allocation policy $\varphi^*$, the embedded Markov chain $\{\mathbf{Q}_m\}$ is ergodic and *time reversible* and the steady state distribution $\Omega^{\varphi^*} = \{\omega^*(q_1,...,q_L)\}$ of the queue length process evolving under the optimal policy $\varphi_i^*$ can be obtained by solving the *L-dimensional* detailed balance equations and the average delay of the $i$th data stream is further given by $\overline{T_i}(\varphi^*) = \sum_{q_1,...,q_L} q_i \omega^*(q_1,...,q_L)$.

As a final step, we shall determine the Lagrange multiplier $\gamma$ by substituting (15) into (4) so as to satisfy the overall average transmit power constraint $P_0$.

$$P_0 = \sum_{q_1,...,q_L} \sum_{i=1}^{L} \mathbb{E}_{\mathbf{H}} \left[ \left( \frac{\delta V_i(\mathbf{Q})}{\gamma \overline{N_i}} - \frac{1}{\alpha(\epsilon)\xi_{[i]}} \right)^+ | \mathbf{Q} \right] \omega^*(\mathbf{Q}) \quad (16)$$

### D. Summary of the Optimal Solution

In this section, we shall summarize the major results derived for delay-optimal performance. The optimal precoding and power allocation policy consists of an online procedure and an offline procedure. They are summarized below.

*Offline Procedure*

- **Determination of Bellman Solution:** For a given $\gamma$, determine $\theta^*(\gamma), \{V^*(q_1,..,q_L;\gamma)\}$ by solving the system of equations according to (14).
- **Transmit Power constraint:** Determine $\gamma$ that satisfies the transmit power constraint in (16).[8]

The outputs of the offline procedure are $\gamma(P_0), \theta^*(\gamma(P_0))$ and $\{\delta V_i^*(q_1,..,q_L)\}$, which shall be used in the online procedure.

*Online Procedure*

- **Step 1) SVD of CSIT:** Given the current CSIT $\mathbf{H}$, obtain the largest $L$ eigenvalues $(\xi_1 \leq \xi_2 \leq ... \leq \xi_L)$ of the matrix $\mathbf{H}^H \mathbf{H}$ and the corresponding eigenvectors.
- **Step 2) Optimal Precoder and Data Stream Index Assignment:** The optimal precoder $\mathbf{P} = \mathbf{U}\Sigma_p$ where $\Sigma_p = diag\{\sqrt{p_1},..,\sqrt{p_L}\}$ and $\mathbf{U} \in \mathbb{C}^{N_t \times L}$ contains the $L$ eigenvectors obtained in Step 1 as columns. The ordering of the $L$ eigenvalues (as well as the corresponding eigenvectors) are sorted in the same order as $\delta V_1^*(\mathbf{Q}),...,\delta V_L^*(\mathbf{Q})$ for the given QSI $\mathbf{Q}$[9].
- **Step 3) Optimal Power Allocation:** Based on the precoder and data stream index association in step 2, the power allocation is given by $\varphi^*(\mathbf{H}, \mathbf{Q}) = P^*(\mathbf{H}, \delta V_1(\mathbf{Q}),..., \delta V_L(\mathbf{Q}))$ as defined in (15).

## IV. LOW COMPLEXITY SOLUTION

While the solution derived in the previous section is optimal and the solution to the Bellman equation (14) can be carried out in an offline manner, the complexity involved is huge as it involves solving for exponentially large (w.r.t. $L$) number of variables (worst case complexity of $\mathcal{O}((N+1)^L)$). In this section, we propose a low complexity suboptimal solution as an alternative, which has a worst case complexity of $\mathcal{O}(NL)$ in the offline procedure but close-to-optimal performance.

### A. Decomposition of the MDP

Using the optimal unitary precoding solution $\mathbf{U}$ in Theorem 1, the Bellman Equation is coupled among the $L$ data streams due to the sorting requirement of the eigenvalues according to $\delta V_1(\mathbf{Q}),..,\delta V_L(\mathbf{Q})$. In order to obtain simple solution, we consider a *static sorting* arrangement for the $L$ largest eigenvalues $\xi_1,..,\xi_L$. Specifically, we shall sort the $L$ eigenvalues in the same ordering as $\beta_1,..,\beta_L$ (which represents the relative importance of the $L$ data streams). While this is suboptimal in strict sense, the proposed *static sorting scheme* will not cause too much performance loss especially for highly asymmetric cases ($\beta_1 \gg \beta_2 \gg ...\beta_L$) or highly symmetric case $\beta_1 \approx \beta_2 \approx ... \approx \beta_L$. Using *static sorting scheme* and given a stationary power control policy, $\varphi = (\varphi_1,..,\varphi_L)$, the MDP state transition probability as depicted in Fig. 2 is decomposable among the $L$ data streams.

---

[8]For simplicity, in our simulation, we avoid using root-finding algorithms to calculate $\gamma$, but calculate the corresponding $P_0$ for each given $\gamma$ by (16).

[9]For example, let $L = 3$. Given the current QSI $\mathbf{Q} = (q_1, q_2, q_3)$, assume $\delta V_1^*(q_1, q_2, q_3) = 2.0, \delta V_2^*(q_1, q_2, q_3) = 3.0, \delta V_3^*(q_1, q_2, q_3) = 1.5$. Then the largest eigenvalue $\xi_{[1]}$ should be associated with the 2nd data stream. The next largest eigenvalue $\xi_{[2]}$ should be associated with the 1st data stream and the smallest eigenvalue $\xi_{[3]}$ should be associated with the 3rd data stream.

The average cost per stage in (6) under $\varphi = (\varphi_1, .., \varphi_L)$ can be decomposed as $J_{\boldsymbol{\beta}}^{\varphi} = \sum_{i=1}^{L} J_{\beta,i}^{\varphi_i}$ where

$$J_{\beta,i}^{\varphi_i} = \lim_{M \to \infty} \frac{1}{M} \sum_{m=1}^{M} g_i(Q_{i,m}, \overline{\varphi}_i(Q_{i,m})), \quad (17)$$

$$g_i(Q_{i,m}, \overline{\varphi}_i(Q_{i,m})) = \beta_i Q_{i,m} + \gamma \overline{\varphi}_i(Q_{i,m)}), \quad (18)$$

$$\overline{\varphi}_i(Q_{i,m}) = \mathbb{E}\left[\varphi_i(\chi_m) | Q_{i,m}\right] \quad (19)$$

Hence, the original "minimal average cost per stage" problem $J_{\boldsymbol{\beta}}^* = \inf_{\varphi} J_{\boldsymbol{\beta}}^{\varphi}$ can be decomposed into $L$ individual sub-problems $J_{\beta,i}^* = \inf_{\varphi_i} J_{\beta,i}^{\varphi_i}$ for $i = 1, .., L$. Consider the $i$-th subproblem, the Bellman equation is given by:

$$\theta_i + V_i(q) = \inf_{\varphi_i(q)} \left\{ g_i(q, \overline{\varphi}_i(q)) + \tau \lambda_i V_i((q+1)_{\bigwedge N}) \right.$$
$$\left. + \tau \overline{\mu_i}(q) V_i([q-1]^+) + V_i(q)(1 - \tau\lambda_i - \tau \overline{\mu_i}(q)) \right\} (20)$$

for all $q \in \{0, 1, .., N\}$, in which $\overline{\mu_i}(q)$ is given by

$$\overline{\mu_i}(q) = \frac{1}{N}\mathbb{E}\left[\log_2(1 + \alpha(\epsilon)\varphi_i(\chi_i)\xi_i) | Q_{i,m} = q\right], \quad (21)$$

where $\xi_i$ is the $i$-th eigenvalue of $\mathbf{H}^H \mathbf{H}$ (sorted in the same order as $\{\beta_1, .., \beta_L\}$) and $\varphi_i(q) = \{p_i = \varphi_i(\mathbf{H}, Q_i = q) : \mathbf{H} \in \mathbb{C}^{N_r \times N_t}\}$ denotes the set of power allocation actions for all CSIT realizations at a given QSI $Q_i = q$. Since the embedded Markov chain $\{Q_{i,m}\}$ is irreducible, there is a unique solution $(\theta_i, V_i(0), ..., V_i(N))$ satisfying (20) and $\theta_i = J_{\beta,i}^*$. We shall derive a low complexity optimal solution for the Bellman equation (20) in the next subsection.

### B. Solution to the decoupled Bellman Equation

Without loss of generality, we shall consider the $i$-th MDP problem. Let $\delta V_i(q) = \tau(V_i(q) - V_i(q-1))$ for $q = 1, 2, .., N$. The Bellman equation in (20) can be expressed recursively in terms of $\{\delta V_i(q)\}$ as follows:

$$\lambda_i \delta V_i(q+1) = \theta_i + \widetilde{\phi}_i(\delta V_i(q)) - \beta_i q \quad (22)$$

for $q = 0, 1, ..., N-1$, with two boundary conditions that $\delta V_i(0) = 0$ and $\beta_i N = \widetilde{\phi}_i(\delta V_i(N)) + \theta_i$, where

$$\widetilde{\phi}_i(y) = \sup_{\{p(\mathbf{H})\}} \mathbb{E}_{\mathbf{H}}\left[\frac{y}{N_i}\log_2\left(1 + \alpha(\epsilon)p(\mathbf{H})\xi_i\right) - \gamma p(\mathbf{H})\right].$$

To solve the Bellman equation in (22), we can first choose a testing value $\theta$ and for each stream and obtain a sequence $\{\delta V_i(1, \theta), ..., \delta V_i(N, \theta)\}$ inductively from (22) for $q = 0, 1, ..., N-1$. Define $f_i(\theta) \triangleq [\widetilde{\phi}_i(\delta V_i(N, \theta)) + \theta]/\beta_i$, the tuple $(\theta, \delta V_i(1, \theta), ..., \delta V_i(N, \theta))$ is a solution to the Bellman equation in (22) if and only if $f_i(\theta) = N$. Since $f_i(\theta)$ is continuous, strictly increasing in $\theta$, there exists a unique $\theta_i^* = f_i^{-1}(N)$ so that $f_i(\theta_i^*) = N$. Correspondingly, $(\theta_i^*, \delta V_i(1, \theta_i^*), ..., \delta V_i(N, \theta_i^*))$ is the unique solution satisfying the Bellman equation in (22) and $\theta_i^*$ can be obtained easily by one-dimensional bisection method. Furthermore, using standard optimization techniques, the optimal power allocation policy (for a given QSI $Q_i = q$ is given by

$$p_i^*(\mathbf{H}, q) = \left(\frac{1}{\widetilde{\gamma}_i} - \frac{1}{\alpha(\epsilon)\xi_i}\right)^+ \quad (23)$$

for $q = 1, 2, ..., N$ and $p_i^*(\mathbf{H}, 0) = 0$.

*Remark 2:* In equation (23), the power allocation solution depends on the QSI only via the equivalent water-level $\widetilde{\gamma}_i^{-1} = \delta V_i(q, \theta_i^*)/\gamma \overline{N}_i$. For larger queue size, the equivalent water-level $\widetilde{\gamma}_i^{-1}$ is increased. This result is also consistent with the asymptotic delay-optimal solution for point-to-point single-stream system in [10].

Using the optimal power allocation policy $\varphi_i^*(q)$ for $q = 0, 1, 2, ..., N$, the embedded Markov chain $\{Q_{i,m}\}$ of the $i$-th data stream is ergodic and time reversible. The steady state distribution $\boldsymbol{\Omega}(\varphi_i^*) = (\omega_0(\varphi_i^*), \omega_1(\varphi_i^*), .., \omega_N(\varphi_i^*))$ of the queue lengths under the optimal policy $\varphi_i^*$ can be obtained by solving the $L$ *one-dimensional* detailed balance equations for all $q = 0, 1, ..., N-1$ combined with $\sum_{q=0}^{N} \omega_q(\varphi_i^*) = 1$.

As a final step for the power allocation policy, we have to determine the common Lagrange multiplier $\gamma$ among the $L$ data streams to satisfy the overall average power constraint

$$P_0 = \sum_{i=1}^{L} \mathbb{E}_{\mathbf{H}}\left[\sum_{q=0}^{N} \omega_q(\varphi_i^*) \left(\frac{\delta V_i(q, \theta_i^*)}{\gamma \overline{N}_i} - \frac{1}{\alpha(\epsilon)\xi_i}\right)^+\right] \quad (24)$$

### C. Summary of the Low Complexity Solution

The low complexity precoding and power allocation policy also consists of an online procedure and an offline procedure, which are summarized below.

*Offline Procedure*

- **Step 1) Determination of Bellman Solutions:** For $i = 1, .., L$ and a $\gamma$, determine $\{\theta_1^*(\gamma), .., \theta_L^*(\gamma)\}$ as well as $\{\delta V_1(q, \theta_1^*(\gamma)), ..., \delta V_L(q, \theta_L^*(\gamma))\}$ according to (22).
- **Step 2) Transmit Power Constraint:** Solve for $\gamma$ that satisfies the transmit power constraint in (24) using one dimensional root-finding numerical algorithm.

The offline complexity is only of $\mathcal{O}(NL)$. The outputs of the offline procedure include $\gamma(P_0), \theta_1^*(\gamma(P_0)), .., \theta_L^*(\gamma(P_0))$ as well as $\{\delta V_1(q, \theta_1^*(\gamma(P_0))), ..., \delta V_L(q, \theta_L^*(\gamma(P_0)))\}$. These shall provide inputs to the online procedure.

*Online Procedure*

- **Step 1) SVD on CSIT:** Given the current CSIT $\mathbf{H}$, obtain the largest $L$ eigenvalues ($\xi_1 \leq \xi_2 \leq ... \leq \xi_L$) of the matrix $\mathbf{H}^H \mathbf{H}$ and the corresponding eigenvectors.
- **Step 2) Precoder and Data Stream Mapping:** The optimal precoder $\mathbf{P} = \mathbf{U}\boldsymbol{\Sigma}_p$ where $\boldsymbol{\Sigma}_p = diag\{\sqrt{p_1}, ..., \sqrt{p_L}\}$ and $\mathbf{U} \in \mathbb{C}^{N_t \times L}$ contains the $L$ eigenvectors obtained in Step 1 as columns. The $L$ largest eigenvalues are sorted in the same order as $\{\beta_1, ..., \beta_L\}$.
- **Step 3) Optimal Power Allocation:** Based on the precoder and data stream index association in step 2, the power allocation of the $i$-th data stream is given by $\varphi_i^*(\mathbf{H}, \mathbf{Q}) = p_i^*(\mathbf{H}, q_i)$ according to (23).

## V. EXTENSIONS TO OUTDATED CSIT

When the CSIT is outdated, there will be spatial interference between the spatial streams of the MIMO channels, which further complicates the precoder design. We shall first define the MIMO physical layer model with CSIT error and extend our delay-optimal formulation and results thereafter.

### A. MIMO Physical Layer Model with CSIT Error

Consider the case where the CSIT error is due to the estimation noise on the reverse link pilot in a TDD scheme, the MMSE estimator of the CSIT $\hat{\mathbf{H}}$ at the transmitter is given by $\hat{\mathbf{H}} = \mathbf{H} + \Delta\mathbf{H}$ [21], where $\Delta\mathbf{H} \sim \mathcal{CN}(0, \sigma_e^2\mathbf{I})$[10]. Moreover, $\mathbb{E}[\Delta\mathbf{H}^H\hat{\mathbf{H}}] = \mathbf{0}$ due to the orthogonality principle of MMSE. Hence, $\sigma_e^2$ is a parameter which represents the CSIT quality[11].

Following similar methods in Section II, we shall extend the MIMO physical layer model to accommodate the effect of the outdated CSIT. Specifically, the conditional average SINR of the i-th stream is given by $\overline{SINR_i}(\mathbf{P}) = \mathbb{E}\left[|\mathbf{w}_i^H\mathbf{H}\mathbf{p}_i|^2/\mathbf{w}_i^H\mathbf{A}_i\mathbf{w}_i \mid \hat{\mathbf{H}}\right]$. Hence, the conditional SER (conditioned on the CSIT $\hat{H}$) of QAM constellation and the associated data rate of the i-th stream $R_i$ are given by $P_e(\hat{\mathbf{H}}) \leq \kappa_1 Q\left(\sqrt{\frac{3\overline{SINR_i}}{2^{R_i}-1}}\right) \leq \frac{\kappa_1}{2}\exp\left(\frac{3\overline{SINR_i}}{2(2^{R_i}-1)}\right)$ and $R_i = \log_2(1 + \alpha(\epsilon)\overline{SINR_i}(\mathbf{P}))$, respectively. Combining the definition in (1) and the matrix inversion lemma [19], we may express the conditional average SINR of the $i$-th stream as

$$\overline{SINR_i}(\mathbf{P}) = \mathbb{E}\left[\mathbf{E}_{ii}^{-1}(\mathbf{P}) - 1|\hat{\mathbf{H}}\right] \geq \mathbb{E}\left[\mathbf{E}_{ii}(\mathbf{P})|\hat{\mathbf{H}}\right]^{-1} - 1$$

where the last step results from Jensen's inequality. Hence, we have a lower bound for the average supported data rate (conditioned on $\hat{\mathbf{H}}$) at the target SER $\epsilon$ given by $R_i \geq \log_2\left(1 + \alpha(\epsilon)(\overline{\mathbf{E}_{ii}}^{-1}(\mathbf{P}) - 1)\right)$, where $\overline{\mathbf{E}_{ii}} = \mathbb{E}\left[\mathbf{E}_{ii}|\hat{\mathbf{H}}\right]$.

### B. Extension of the Formulation and Results

The delay optimization problem formulation in (5) can be easily extended for outdated CSIT by modifying the system state variable $\chi = (\hat{\mathbf{H}}, \mathbf{Q})$. Theorem 1 can be extended as

*Corollary 1:* For any realization of system state $\chi(\hat{\mathbf{H}}, \mathbf{Q})$, the optimal precoding action $\pi(\chi) = \mathbf{P}$ w.r.t. (10) is given by:

$$\pi(\chi) = \mathbf{P} = \mathbf{U}\boldsymbol{\Sigma}_p \qquad (25)$$

where $\mathbf{U} \in \mathbb{C}^{N_t \times L}$ is a unitary matrix consisting of $L$ eigenvectors of $\hat{\mathbf{H}}^H\hat{\mathbf{H}} + N_r\mathbf{I}$ corresponding to the $L$ largest eigenvalues and $\boldsymbol{\Sigma}_p = diag\{\sqrt{p_1}, \ldots, \sqrt{p_L}\}$ is a diagonal matrix containing the power allocations over the $L$ spatial channels. Note that the $L$ largest eigenvalues $\{\xi_1, .., \xi_L\}$ are sorted in the same order as $\eta_i = V(q_1, \ldots, q_L) - V(q_1, \ldots, [q_i - 1]^+, \ldots, q_L)$.

Using the precoder structure given by Corollary 1, the conditional average MSE becomes $\mathbf{d}[\overline{\mathbf{E}}] = [(1+p_1\xi_1)^{-1}, \ldots, (1+p_L\xi_L)^{-1}]$ [19]: and hence, the conditional average service rate $\overline{\mu}_i$ becomes $\overline{\mu_i}(\mathbf{Q}) = \mathbb{E}_{\hat{\mathbf{H}}}\left[\frac{1}{N_t}\log_2(1 + \alpha(\epsilon)p_i\xi_i)\right]$. As a result, all the subsequent formulation and solutions can be applied by replacing the $\mathbf{H}$ with $\hat{\mathbf{H}}$ as the estimated CSIT.

---

[10]For detailed error model, please refer to our full version [20]

[11]We assume that the receiver has perfect knowledge of CSIR for detection and decoding. This is because that in practice, a relatively strong forward link pilot channel is available from the base station to the receivers, so that the CSIR estimation error is insignificant relative to that of the CSIT.

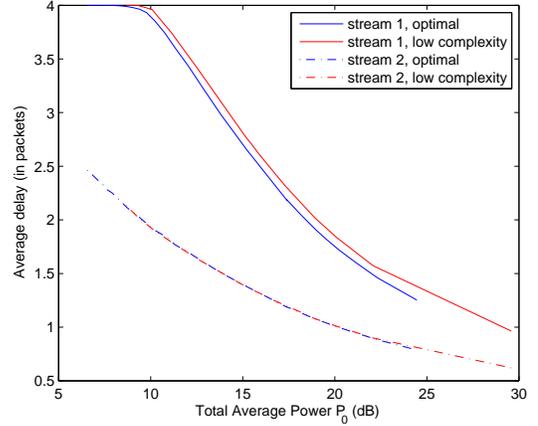

Fig. 3. Comparison of the average delay under optimal and low complexity solutions under perfect CSIT. $N_t = N_r = 2$, $\beta_1 = 1$, $\beta_2 = 10$.

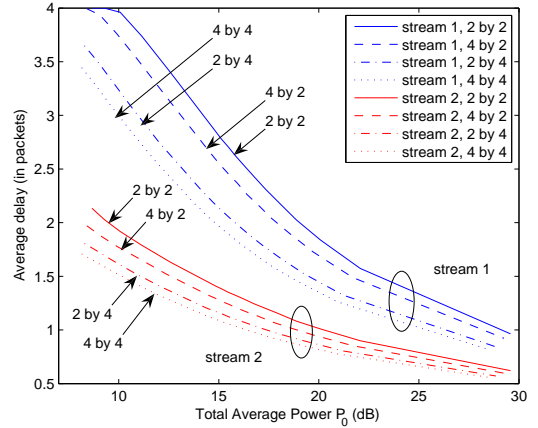

Fig. 4. Average delay of the proposed low-complexity solution for different $(N_t, N_r)$ configurations under perfect CSIT. $\beta_1 = 1$, $\beta_2 = 10$.

## VI. NUMERICAL RESULTS AND DISCUSSIONS

In this section, we evaluate the proposed solutions to the delay sensitive precoder and power adaptation design via numerical simulations. Two data streams are considered with weights $\beta_1$, $\beta_2$ in (5), respectively. The mean packet size and mean arrival rate for the two streams are the same, i.e. $\overline{N_1} = \overline{N_2} = 200$ bits per packet and $\lambda_1 = \lambda_2 = 0.02$ packets per channel use time $\tau$. The buffer size is $N = 4$ for each stream[12]. The scheduling time unit $\tau$ and the target SER $\epsilon$ are fixed at 5ms and 1%, respectively.

Fig. 3 compares the average delay of the two data streams under the optimal and low complexity solutions for a 2-by-2 MIMO system. As we can see from the figure, both of our proposed solutions show full support of heterogeneous delay-sensitive services. Furthermore, the low complexity solution has close-to-optimal performance with a worst case complexity of only $\mathcal{O}(NL)$, which indicates its practical significance.

Fig. 4 depicts the average delay of the two streams of the proposed low-complexity solution under different configurations of transmit and receive antennas. In Fig. 5, we set $\beta_1 = \beta_2 = 1$ and compare the sum average delay of the proposed scheme for a 2-by-2 MIMO system with two baselines: 1) the *Round-Robin scheme*, i.e. the two streams are serviced in TDMA fashion with equally allocated time slots; 2) the *CSIT only scheme*, i.e. the precoder and power adaptation for the two streams are designed purely based on the outdated CSIT. Above 10dB gain can be achieved by the proposed scheme over the two baselines. The figure also suggests that spatial multiplexing may not help effectively without adapting to both the CSI and the QSI, and the *CSIT only scheme* is much more sensitive than the proposed scheme w.r.t. the CSIT quality. This illustrates the robustness of our proposed scheme to CSIT errors.

---

[12]This implies that the delay for a packet is at most four packets. Since we are considering delay-sensitive applications, this can be a valid assumption.

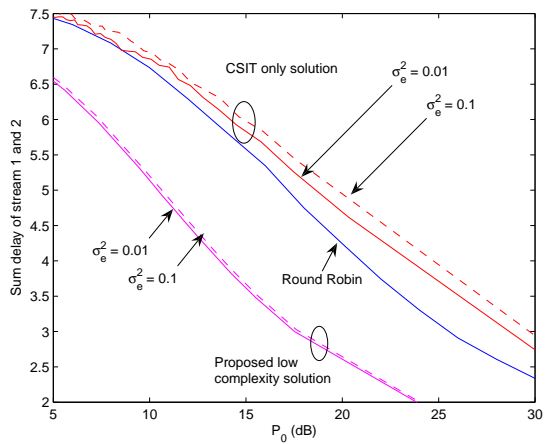

Fig. 5. Sum average delay of the proposed low complexity scheme and two baseline schemes, given different values of CSIT error variance $\sigma_e^2$, $\beta_1 = \beta_2 = 1$, $N_t = N_r = 2$.

## VII. Summary

We considered delay sensitive MIMO system with $L$ heterogeneous data streams spatially multiplexed together. The design of precoding policy achieving Pareto optimal delay tradeoff is fomulated into an $L$-dimensional MDP problem. A low complexity solution with worst case complexity $\mathcal{O}(NL)$ is proposed by decomposing the original problem into $L$ one-dimensional subproblems based on static sorting. Numerical results verify the advantages of taking both QSI and CSIT error into dynamic precoder design.

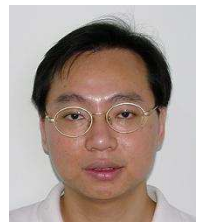

**Vincent Lau** obtained B.Eng (Distinction 1st Hons) from the University of Hong Kong (1989-1992) and Ph.D. from Cambridge University (1995-1997). He was with HK Telecom (PCCW) as system engineer from 1992-1995 and Bell Labs - Lucent Technologies as member of technical staff from 1997-2003. He then joined the Department of ECE, Hong Kong University of Science and Technology (HKUST) as Associate Professor. His current research interests include the robust and delay-sensitive cross-layer scheduling of MIMO/OFDM wireless systems with imperfect channel state information, cooperative and cognitive communications, dynamic spectrum access as well as stochastic approximation and Markov Decision Process.

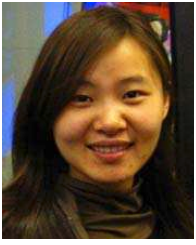

**Yan Chen** received her B.Sc degree from Chu Kochen Honored College, Zhejiang University, Hangzhou, China, in 2004. She is expected to receive her Ph.D degree in information and communication engineering from Zhejiang University in 2009. Since Jan 2008, She has been a visiting researcher in the group of Prof. Vincent Lau in HKUST. Her current research interests lie in combined information theory and queueing theory in wireless communications, with particular emphasis on exploiting communication opportunities via cooperation and cognition.

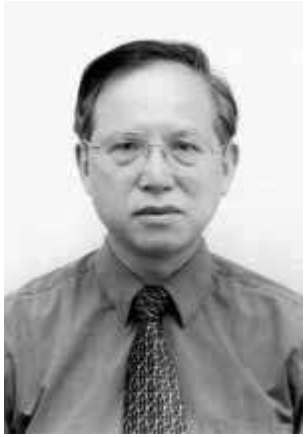

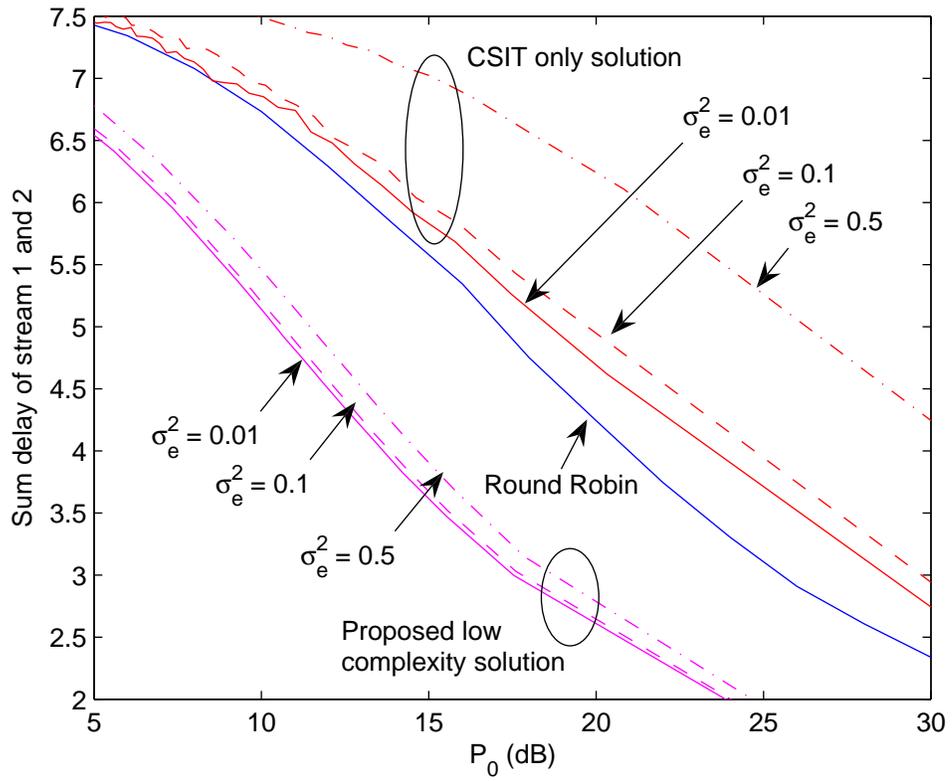

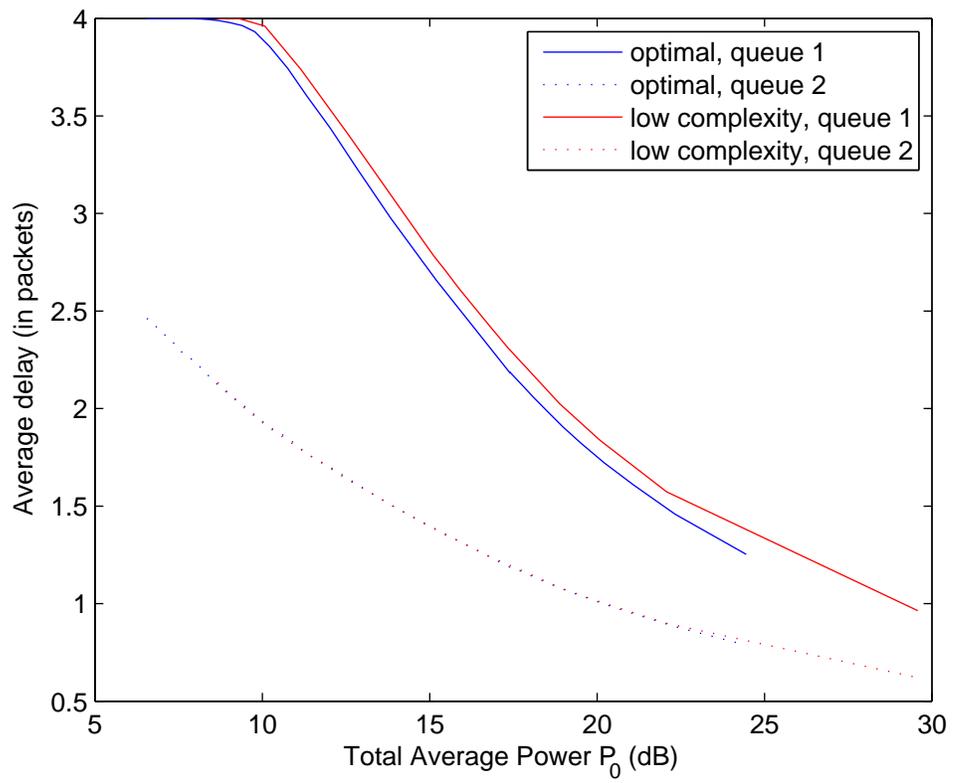

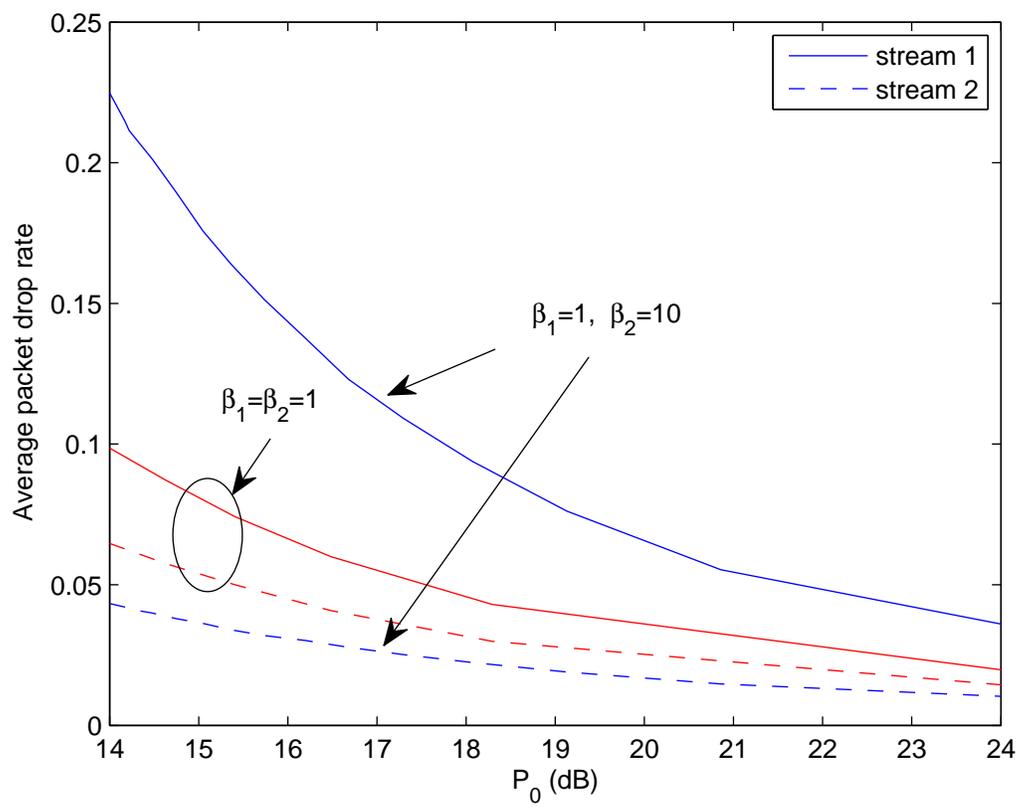

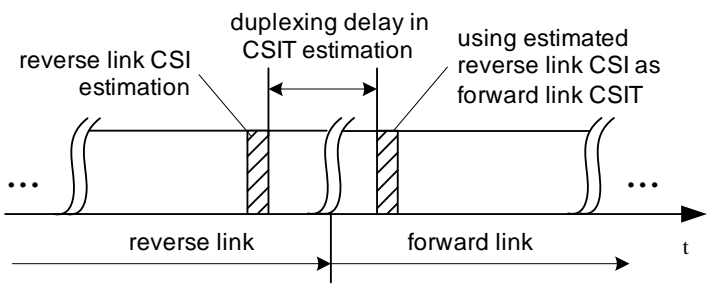

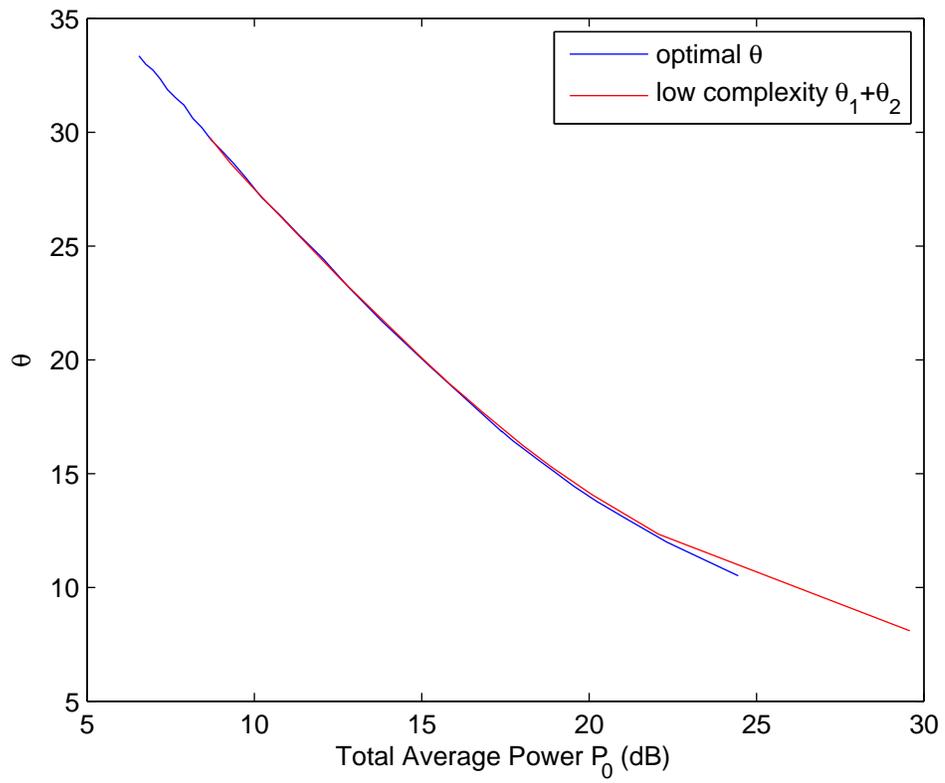